\documentclass[sn-mathphys,Numbered]{sn-jnl}

\usepackage{graphicx}
\usepackage{appendix}
\usepackage{url}
\usepackage{verbatim}
\usepackage{color}
\usepackage{colortbl}
\usepackage{xcolor}
\definecolor{lgray}{gray}{0.97}
\usepackage{mdframed}
\usepackage{tikz}
\usetikzlibrary{positioning}
\usetikzlibrary{shapes.geometric,backgrounds,calc}
\usetikzlibrary{fit}
\usepackage{amsmath}

\usepackage{perpage}
\usepackage[perpage]{manyfoot}

\newcommand{\ie}{\emph{i.e.}, }                                                         
\newcommand{\eg}{\emph{e.g.}, }

\begin{document}

\title[Modeling Rabbit-Holes on YouTube]{Modeling Rabbit-Holes on YouTube}

\author[1]{\fnm{Erwan} \sur{Le Merrer}}\email{erwan.le-merrer@inria.fr}

\author[2]{\fnm{Gilles} \sur{Tredan}}\email{gtredan@laas.fr}

\author[1]{\fnm{Ali} \sur{Yesilkanat}}\email{ayesilk@inria.fr}

\affil[1]{\orgname{Inria}, \orgaddress{\city{Rennes}, \country{France}}}

\affil[2]{\orgdiv{LAAS}, \orgname{CNRS}, \orgaddress{\city{Toulouse}, \country{France}}}

\abstract{
Numerous discussions have advocated the presence of a so called \textit{rabbit-hole} (RH) phenomenon on social media, interested in advanced personalization to their users. This phenomenon is loosely understood as a collapse of mainstream recommendations, in favor of ultra personalized ones that lock users into narrow and specialized feeds. Yet quantitative studies are often ignoring personalization, are of limited scale, and rely on manual tagging to track this collapse. This precludes a precise understanding of the phenomenon based on reproducible observations, and thus the continuous audits of platforms.\\
In this paper, we first tackle the scale issue by proposing a user-sided \textit{bot}-centric approach that enables large scale data collection, through autoplay walks on recommendations. We then propose a simple theory that explains the appearance of these RHs.  While this theory is a simplifying viewpoint on a complex and planet-wide phenomenon, it carries multiple advantages: it can be analytically modeled, and provides a general yet rigorous definition of RHs. We define them as an interplay between \textit{i)} user interaction with personalization and \textit{ii)} the attraction strength of certain video categories, which cause users to quickly step apart of mainstream recommendations made to fresh user profiles.\\
We illustrate these concepts by highlighting some RHs found after collecting more than 16 million personalized recommendations on YouTube. A final validation step compares our automatically-identified RHs against manually-identified RHs from a previous research work. Together, those results pave the way for large scale and automated audits of the RH effect in recommendation systems.
}

\keywords{Recommendation, rabbit-holes, modeling, data-collection}

\maketitle

\section{Introduction}

Modern recommendation systems, and in particular personalization on video platforms, are tuned to maximize users watch time \cite{youtube} and their presence on platforms. This has been associated with major societal impacts \cite{stray2021you}, triggering a debate on whether or not the personalization brings users on \textit{radicalization paths} \cite{radicalization, ledwich2019algorithmic}. To a less extreme end, the problem of users getting a narrow view on the possibilities offered by the platforms has been studied qualitatively under the terms of \textit{filter bubbles} \cite{pariser2011filter,recsys-bubble, whittaker2021recommender}, and more recently the \textit{rabbit- hole} (shorthanded RH through the paper) phenomenon \cite{ledwich2019algorithmic,kaiser2019implications,o2015down,sashittal2021brands}.

The dynamics of the personalization by these recommendation systems is based on opaque factors, and their operating platforms appear as black boxes to their users \cite{pasquale2015black}. 
Yet the current understanding is that the RH effect can be defined as the situation where "recommendations influence viewers of radical content to watch more similar content than they would otherwise" \cite{ledwich2019algorithmic}. This stresses a dynamic where the watch history of users is a crucial preamble to future recommendations, but also that there is an average case where these users would not access these videos otherwise (\ie most probably would view more mainstream content). Such definitions are certainly hard to leverage when it comes to practical quantification, for instance when auditing such a platform. While tools for measuring simpler phenomena such as popularity of videos with regards to geography \cite{popularity}, or the amount of ads \cite{ads} for instance emerged, such tools are lacking for quantifying the RH effect. Despite its importance for the rising field of algorithm auditing \cite{raji2020closing}, quantification is inherently a challenge, as it requires to quantify "average" experiences from a collection of experiences made unique by the very process of personalization.
This paper thus seeks to articulate a precise definition of the RH effect so that reproducible and automated means of measure are made possible (rather than relying on manual tagging \cite{radicalization,ledwich2019algorithmic}). 

We take the step to collect personalization to multiple and controlled user profiles in YouTube, in order to match a theory we propose under the form of a model. This model is composed by the two core components at stake in the previous definition \cite{ledwich2019algorithmic}: there is \textit{i)} a \textit{feedback loop} in the viewing patterns of users, for \textit{ii)} specific types of videos that have the characteristic to be far away from a "standard" usage. In this paper we coin these videos \textit{attractors}, and we will characterize them by their capability to collapse the entropy of more mainstream video recommendations. The feedback loops for personalization to a user will be modeled and measured by using the \textit{autoplay}-recommendations \cite{masadeh2020aftermath,stocker2020riding,10.1145/3411764.3445467} provided by YouTube to that user.

\paragraph{Contributions} This paper makes the following main contributions. 
\textit{i)} We present in Section \ref{personalization} an experimental setup that allows to \textbf{measure recommender personalization through bots} and validate this experimental setup on a large collected corpus.
This is key to automate the study of the trapping dynamics of users into RHs. 
\textit{ii)} We propose a simple \textbf{model to explain the formation of RHs}. This model permits to detect users entering RHs by relative comparison (Section \ref{model}). 
Finally, \textit{iii)} we validate our model on the personalized data collected by bots, quantify the attractivity of some channel categories, show that we can \textbf{automatically identify the RHs} that were manually labeled by a previous research work, and discuss a \textbf{measure of the personalization strength} of any video.

\section{Background}

\paragraph{The lack for a precise understanding of the filter bubble or RH effect}
Defining a RH is no easy task. In \cite{ledwich2019algorithmic}, a RH is implicitly defined as filter bubbles where: "recommendations influence viewers of radical content to watch more similar content than they would
otherwise". In \cite{kaiser2019implications}, the authors (although not precisely defining it so) relate their finding to communities in the network formed by recommended channels. Both papers point to work in \cite{o2015down}, which relies on clustering (precisely, non-negative matrix factorization) to study how "\textit{recommendation} can have the undesirable consequence of a
user being excluded from information that is not aligned with their existing perspective, potentially leading to immersion within an ideological bubble".  
All above references point to Pariser's definition of filter bubbles \cite{pariser2011filter}: "Together, [recommendation engines] create a unique universe of information
for each of us—what I’ve come to call a filter bubble—which fundamentally alters the
way we encounter ideas and information".
Interestingly, Pariser insists in the individualized nature of this bubble ("You're alone in it"), while more recent works rather focus on the social effect of such bubbles (\eg considered as a pathway to conspiracy theories). These rather insist on the dangers of community wide bubbles that would attract many users into specific recommendation regimes. 
While these seminal works all refer to the same phenomenon, by defining RHs through their consequences, they leverage subjective notions (ideology, radicalization) that are hard to quantify. Yet, we believe that a rigorous quantification of the RH phenomenon is essential for continuous and automated audits.

\paragraph{Personalization is key to the phenomenon, yet difficult to study}
\textit{Personalization} leverages both the user information and content information to provide tailored recommendations \cite{li2010contextual}. This is obviously the core mechanism to study when one wants to assess the impact of a recommendation algorithm onto users consumption. Unsurprisingly though, personalization is complex to collect and study at scale, as multiple profiles must be tracked. This is why state of the art papers on understanding filter bubbles or RHs solely consider \textit{contextual} recommendation, which is not adapted to users but a generic proposal related to  consumed items \cite{radicalization,papadamou2020understanding}. A recent approach by Ribeiro et al. \cite{radicalization} for instance collects the channels recommended by other selected channels, as general and static contextual references. Roth et al. collect contextual recommendations around certain seed video categories \cite{roth2020tubes}. In \cite{ledwich2019algorithmic}, recommendations made to a single anonymous profile are collected; this profile does not simulate the watching of any video, precluding personalization; human labeling of videos is also performed.
Hussein et al. \cite{misinf} nevertheless managed to setup a collection by bots (through the scriptable Selenium browser), yet leveraged human annotators to label videos (debunking, neutral, promoting) for a study on misinformation prevalent in YouTube searches and recommendations.
These works clearly establish the existence of RHs in contextual recommendations; however, we here argue that personalization is a key element to consider and study in order to explain the current experience of users with regards to RHs.

\section{Surfing personalization on YouTube using bots}
\label{personalization}

\subsection{Bots for collecting personalization at scale}
\label{botperso}

We rely on the use automated scripts (\ie \textit{bots}) in order to simulate human users watching and interacting with videos on YouTube. 
Our bot actions are implemented using the scriptable Puppeteer headless Chrome browser\footnote{\url{https://github.com/puppeteer/puppeteer} -- We note that recommendations (be they contextual or personalized) are only available at the time of our collection campaign using a Javascript enabled browser: the mere use of a fast yet basic static web scrapper (such as \eg scrappy) does not allow recommendation collection, as it was the case several years ago \cite{le2017topological}.}. The most important actions we have implemented for this study are: the search for a given term in the search bar, the mouse click on a recommendation based on its position in the list, the collection of videos from a given user channel, the viewing of a video (and the associated collection of recommendations), and the access to the YouTube welcome page to gather context-free recommendations before/after searches and video views by a bot.

To control personalization as much as possible, and avoid measurement artifacts due to browser fingerprinting for instance, we took the following steps.
The bots are built within Docker containers that are instantiated and then destroyed after each user simulated interaction with YouTube, so that we are sure to remove all traces and variables between multiple runs. This is in particular the case for cookies, that are thus reinitialized every time.
At the run of a new bot for a particular action and data collection, we randomize variables such as the browser user agent, and announced screen resolution. The watch time for videos by the bots is randomized to a mean of $5$ minutes $\pm 10\%$ (if videos last at least this duration of course)\footnote{This value is way over the 30 seconds known to be the limit of YouTube to consider a real video watch and then to increment its view counter, see \eg  {\url{https://growtraffic.com/blog/2017/08/youtube-video-counts-view}}.}.

We summarize and depict the instrumentation of this study in Appendix Section \ref{appendix:instrumentation}. We report a lightweight operational execution of each bot, that permits its execution on basic machines such as on the four Intel NUC computer architectures we used in parallel for this study.

\paragraph{Data collection summary}
Our automated data collection campaign took place from February 18th to March 22nd, 2022.
The recommendations were captured on the YouTube welcome page, by a scroll-down fetch resulting in around $300$ recommendations gathered each time, before and after each video watched by a bot. 

We summarize these statistics in Table \ref{table:data}.
For the experiments concerning our contribution \textit{i)}, bots are viewing in sequence $20$ videos from kid channels, or from the welcome page as a control experiment. These walks collected personalized recommendations, that were in turn automatically labeled by the tool we will introduce in the sequel. We note that each walk duration is in the order of $100$ minutes, since bots "watch" $5$ minutes of each video along their $20$ hops walk.
This first experiment will reveal that 3 videos in a row are enough to trigger strong personalization (Figure \ref{figexp1}); for scaling to a more massive collection in the next experiment, our bots walks are thus shortened to $5$ videos in a row: in contribution \textit{ii), bots started} from various channels pertaining to several well separated categories that will be discussed (and taken from \cite{radicalization}, and from famous channels for kids), and from a neutral profile as a control experiment. Walks duration are then of length around $25$ minutes. This sums up to $16,386,099$ personalized recommendations collected.

\begin{table}[]
\centering
  \tiny{
\begin{tabular}{ccc}
\hline
& Personalization to bots (Sec. \ref{ss:growing}) & RHs vs channel categories (Sec. \ref{validation}) \\ 
\hline
launched bots (walks) & $533$  & $5,438$ \\ 
length of walks (hops) & 20  & 5 \\
recom. collected  &  $5,746,494$  & $10,639,605$  \\ 
auto. labeling resolved & $5,726,296$ & N/A \\ 
\hline
\end{tabular}
}
\caption{Statistics on the collected data. Bots viewing videos and scraping the resulting YouTube personalization.
} \label{table:data}
\end{table}

\paragraph{From recommendations to similarities between users}

This study relies on bots following so called \textit{Autoplay recommendations} (as studied in related works \eg \cite{masadeh2020aftermath,stocker2020riding,10.1145/3411764.3445467}). When watching a video, the corresponding autoplay recommendation is the top recommendation displayed (right hand side of the YouTube video page), and is the next video to be played if no other action is taken by the user. It certainly indicates a high confidence of the recommendation algorithm that it is suited to the viewer. 

The principal substrate we are thus building on for this study is personalization to simulated user profiles, under a \textit{vector} representation. This vector space corresponds to the \textit{one-hot encoding} of each of the encountered video ID.
For each profile, we thus associate a vector keyed by video IDs (\eg \texttt{A5F9jOHbi5I}), containing the integer representing the amount of times this particular video has been recommended to the bot at a given step (0 if never)\footnote{Hence, we leave aside the position of a given video in the recommendation list. Although this definitely constitutes a valuable information, weighting the relative importance of videos being recommended first or last is difficult and left to future work.}. We call these \emph{recommendation vectors}.
All the vectors corresponding to our collection campaign are then perfect candidates to compute similarities between each of then, using for instance the \textit{cosine similarity} (hereafter denoted $S_c$), well established in the field of recommender systems \cite{cosine-recom}. 

While bots collecting personalization and these vectors are sufficient to study the RH effect, we first introduce the labeler we leverage to validate the instrumentation approach.

\paragraph{An automatic labeler for testing personalization to bots: labeling videos using YouTube Kids}

A first salient question is whether or not a fully automated process as we depicted yields a reaction (\ie personalization) by the YouTube recommendation algorithm. 

For this validation step, we rely on an automated way to label collected recommendations (as opposed to human tagging in \eg \cite{radicalization, ledwich2019algorithmic, misinf}), in order to observe if YouTube recommendations are increasingly composed of the category of videos viewed by the bots. If so, YouTube reacts to the bots actions as expected, and thus automation is a way to observe recommendations based on input actions that we can script. 

YouTube Kids\footnote{\url{https://www.youtubekids.com}} is the alternative for kids to YouTube \cite{bringing-kids}, targeting the following profiles:  "preschool" are 4 years old or less, "younger" are from 5 to 7, and "older" are from 8 to 12. We will generally call a \textit{kid} user one that belongs to one of these three categories. Yet, while YouTube is encouraging kids to use YouTube Kids instead, all major content creators for kids (counting hundreds of millions of views, as reported on Table \ref{table:channels}) are present on YouTube.

We found out that videos share the same video ID on both YouTube and YouTube Kids. Moreover, videos provided by YouTube Kids contain an age tag in the associated page metadata. We leverage this information as an age tracker in the following way. Using the ID of any video encountered in YouTube recommendations, we query YouTube Kids. As a result, either we obtain the video page from YouTube Kids, in which case we know it is a "kid" video and we can extract its precise age label from the page tag, either we obtain an error page, in which case we label it an "adult video" since YouTube did not make this video available for kids. We report that only $533$ videos out of $5,726,296$ failed to be resolved in our dataset.

Using such a labeler for automatic labeling is convenient, as opposed to the tedious task of manual labeling; we now leverage it for observing the personalization proposed to our bots.

\subsection{Measuring personalization on kids profiles} 
\label{ss:growing}

To validate our YouTube instrumentation, we measure the ability of our bots to trigger personalized recommendations that depend on their past video consumption. Thanks to the labeler, we can test the following hypothesis: "bots watching children content will get an increase in recommended children content".

Our approach is as follows: given a list of famous children video channels (that we list in Table \ref{table:channels}), we emulate fresh profiles that watch only videos from these channels. After each video watched, we collect recommendations available on the bot welcome page. For comparison, a control experiment is realized by fresh profiles that watch random videos picked from their welcome page before any action is performed. All recommendations collected by both random and kid profiles are then automatically labeled.
More specifically, we perform multiple runs where a bot starts from one of the five most accessed kids channels on YouTube, and  watches in sequence 20 randomly selected videos from that channel. 

\begin{figure}[h!]
\centerline{\includegraphics[width=.9\linewidth]{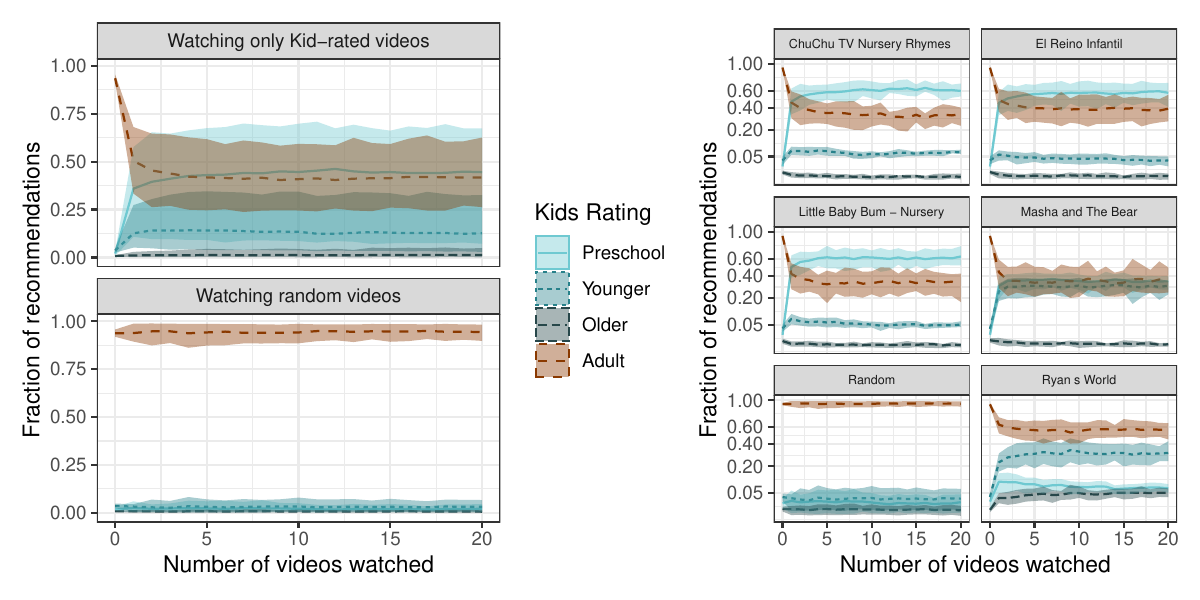}}
\caption{The impact of bots actions on personalization: (\textit{left}) Evolution of the fraction of recommended kid rated videos, when watching either random or kid videos. (\textit{Right}) A detailed view by kid channels (random baseline included for reference). The shaded areas delimit the 10 and 90 percentile.}
\label{figexp1}
\end{figure}

Figure~\ref{figexp1} presents the results of this experiment, for which $5,726,296$ recommendations were collected ($51.3\%$ from random profiles, $48.7\%$ from kids profiles). The figure on the left presents the evolution of the recommendations composition (on the $y$-axis) for the kid profiles (top-figure) and the random baseline (bottom-figure) as a function of the number of watched videos ($x$-axis). We note that originally ($x=0$), the watch history of all profiles being empty, videos for the two profiles get the same recommendation mix containing on average $93.7\%$ of adult (\ie non kid), $2\%$ preschool, $1.8\%$ younger and $0.2\%$ older videos. 
This proportion remains unchanged when watching random videos, which is the expected behavior of our control condition. At contrary, recommendations to kid profiles quickly adapt to converge to a stable mix containing only $41.5\%$ of adult videos and $44.7\%$ preschool, $12.7\%$ younger and $1.2\%$ older. 
In other words, as compared to the baseline, the ratio of non adult videos has been multiplied by  more than 9.2. We draw two main conclusions from this experiment: \emph{i)} Our experimental setup effectively manages to trigger personalization from the recommendation algorithm, and \emph{ii)} this personalization is triggered quickly: the recommendation mix changes drastically within the first 3 watched videos, and further consumed videos have little to no impact.

Figure~\ref{figexp1}-(right) is identically constructed, but provides a detailed view at each kid channel profile. Note the square-rooted $y$-axis scale. While this view confirms the previous conclusions, one can further observe that different kid channels trigger different personalizations. Channels providing preschool content (\eg Little Baby Bum) have the highest impact on diminishing adult recommendations, while channels targeting older kids (\eg Ryan's World) trigger less preschool recommendations. Interestingly, no channel manages to substantially increase the amount of older-type recommendations. While providing a definite explanation would require deeper investigations, we believe this is mostly due to older-type videos being quite rare compared to other categories for kids, perhaps because it is difficult to draw a separation line between older-type and adult-type videos.

\begin{mdframed}[backgroundcolor=lgray]
\textbf{Summary:}  Our bot instrumentation manages to trigger YouTube personalization. This personalization, in the case of kid profiles, quickly kicks in, and impacts a significant fraction of the provided recommendations.
\end{mdframed}

\begin{table}[]
\centering
\tiny{
\begin{tabular}{llll}
\hline
Channel type & Channel name &  \#views  & \#videos \\ \hline
\rowcolor[HTML]{EFEFEF} Kids (K) & Ryan's World &      48,975,229,688     &  2,091  \\
\rowcolor[HTML]{EFEFEF} Kids (K) &    El Reino Infantil       &     44,074,006,603  & 1,157  \\
\rowcolor[HTML]{EFEFEF} Kids (K) &  ChuChu TV Nursery Rhymes \& Kids Songs &  35,098,632,207 & 452 \\ 
\rowcolor[HTML]{EFEFEF} Kids (K) & Little Baby Bum - Nursery Rhymes \& Kids Songs &    34,570,591,653    & 1,669  \\ 
\rowcolor[HTML]{EFEFEF} Kids (K) &   Masha and the bear &  17,643,829,073     &  857  \\ 
Alt-Right (AR) & Black Pigeon Speaks  & 40,774,438    &  175        \\
Alt-Right (AR)  & The Golden One  &  12,286,875    &  605       \\ 
Alt-Right (AR)   & NeatoBurrito Productions  & 8,334,309    &   320     \\ 
Alt-Right (AR)    &  Australian Realist &   5,472,094  &     290     \\
Alt-Right (AR)     & Prince of Zimbabwe  &  40,088,933   &  20       \\ 
Alt-light (AL)      &  StevenCrowder & 1,443,559,041    &    1,296     \\ 
Alt-light (AL)       & Rebel News  & 628,496,085 &  16,535      \\ 
Alt-light (AL)        &  Paul Joseph Watson & 517,887,497    &    829     \\ 
Alt-light (AL)         & Mark Dice  & 440,494,445    &    1,522      \\
Alt-light (AL)          & Sargon of Akkad  &  306,059,074   &  1,187      \\ 
Intellectual dark web (IDW)            & PowerfulJRE  &  1,704,287,921    &    600    \\
Intellectual dark web (IDW)             & JRE Clips  & 4,421,584,838    &     5,393    \\
Intellectual dark web (IDW)              & PragerU  & 1,439,923,494   &     1,755    \\ 
Intellectual dark web (IDW)               & The daily Wire  &  763,331,055   &  6,940      \\ 
Intellectual dark web (IDW)                & The Rubin Report  &  419,351,645    & 3,109    \\ 
Media (M)  &  Vox &  2,734,974,347   &    1,357     \\
Media (M)                  & GQ  & 2,565,984,292    &  1,846     \\
Media (M)                   & VICE News  & 2,364,198,449    &    5,843    \\ 
Media (M)                    & WIRED  & 2,565,984,292 &   3,308     \\ 
Media (M)  & Vanity Fair   & 1,711,181,309    &   2,430  \\ 
\hline
\end{tabular}
}
\caption{Statistics on the channels used as video pools and autoplay walk starts in this study. In this paper we bring the channels for kids (grey rows); these are selected after a top-view ranking from \url{https://www.statista.com/statistics/785626/most-popular-youtube-children-channels-ranked-by-subscribers/}. The four remaining categories are taken from Ribeiro et al.~\cite{radicalization}.}
\label{table:channels}
\end{table}

\section{A model for the Rabbit-Hole effect}
\label{model}
In the previous section, we observed the impact of watched videos (\ie history) on recommendations. Since this observed personalization is the 'raison d'être' of many recommenders, it does not alone justify the existence of RHs.
In this section, we thus investigate the trapping dynamics of users into RHs. More precisely, we elaborate a minimalist model that emulates them. Through this model, we argue that the formation of RHs relies on two ingredients: $i)$ the user interplay with personalization, and  $ii)$ the presence of video sets we call \emph{attractors}.

\subsection{Feedback loops and attractors}

\paragraph{A simple model} Let us first consider a recommender $R$. Suppose the set of videos is containing only two types of videos: $A$ and $B$-types. $R$ works as follows: for every user, $R$ remembers the 10 last  watched videos. When issuing recommendations, $R$ recommends a $B$-type video with a probability $p_B$ proportional to the number of $B$-type videos already in history. For instance, if a user has $5/10$ $B$-type videos in her history, recommendations will be of type $B$ with probability $p_B=1/2$. Once $R$ has selected the video type ($A$ or $B$), it picks the video to recommend uniformly at random among videos of that type. 
$R$ thus captures the following simple intended property: imagine $A$ are Adult videos, and $B$ kid videos; the more kid videos are watched, the more kid videos are recommended.

Users always follow recommendations by watching a uniformly selected random video among all recommended ones\footnote{We note that the prevalence of clicks on recommendations over searches is major, around $70\%$ in 2018 as indicated by YouTube product chief, please refer to \url{https://www.cnet.com/news/youtube-ces-2018-neal-mohan/}.}. This essential component of the model creates a feedback loop between $R$ and the user: the more she watches $B$ the more she gets, which, if she follows recommendations, will get her to watch more $B$ videos and so on and so forth. 

To sum up, this model has essentially two key factors: \textit{i)} a \emph{recommender} $R$, which users always leverage by following recommendations, closing the \emph{feedback loop} between recommendation and history: together, $R$ and users form a dynamical system.
\textit{ii)} A strong distinction between video types: videos types are either $A$-type, either $B$-type. Moreover, this type distinction is associated with a straightforward reaction from $R$: consumption of $B$-type videos systematically increase the likelihood of $B$-type recommendation. As a reference to their key role in the recommendation dynamical system, we refer to $B$ as an \emph{attractor}.
This resulting model is a discrete dynamical system that we can  analyze and simulate. Please refer to Appendix \ref{appendix:model} 
for a more formal description of the model. To maintain a high level perspective, we now focus on simulation.

\begin{figure}[h!]
\centerline{\includegraphics[width=.99\linewidth]{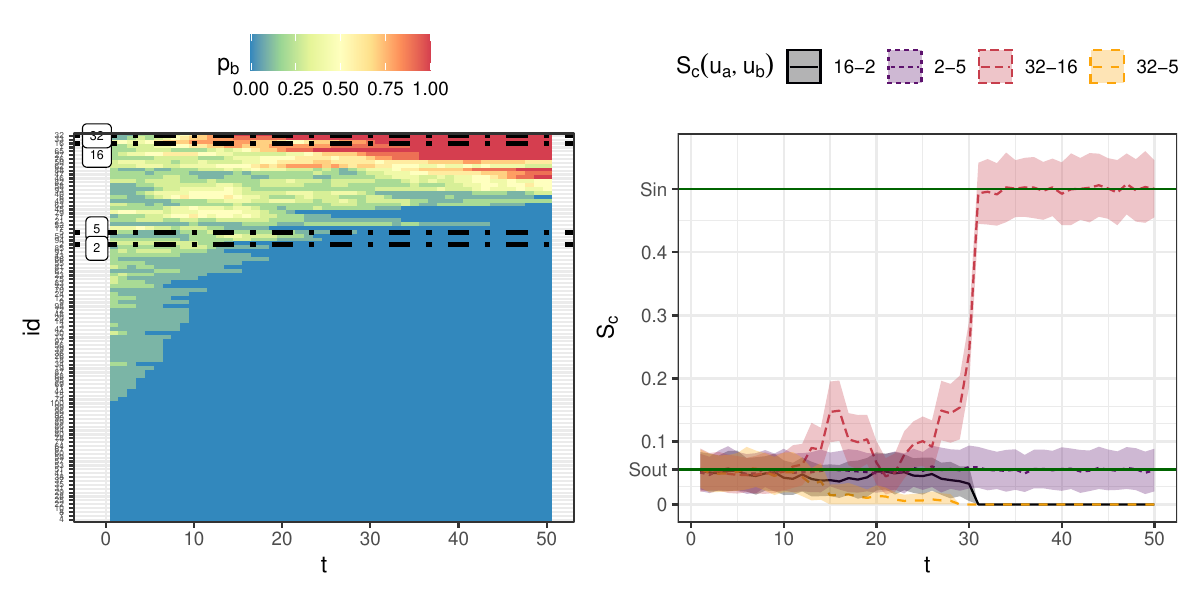}}
\caption{Simulations of the model: \emph{(left)} Evolution of $p_b$ along rounds ($x$ axis) for all users (ordered by average $p_b$, $y$-axis). When $p_b=0$ (blue), users never get recommended any video from $B$ (the RH); when $p_b=1$ (red), users always get recommended $B$ videos. Dashed lines highlight users showcased in the Figure on the right. \emph{(right)} Recommendation (cosine) similarities for four particular users: divergence emerges between a user in the RH (user 32) and a user getting mainstream recommendations (user 5). Recommendation similarity is high between both users in the RH (32 and 16). Value $Sin$ (resp. $Sout$) represents the theoretical similarity value, given by the model, between two users in (resp. out) of the RH.}
\label{toymodel}
\end{figure}

\paragraph{Simulations using the RH model}
We simulate a system of $n=100$ users iteratively obtaining $y=50$ recommendations among a catalog $V$ of size $|V|=1000$ videos. We consider such a set of $|B|=100$ specific videos ($|A|=900$ being in the general category). Users' history is of size $h=10$ and is initially filled with videos picked uniformly randomly (\ie belong to $B$ with probability $|V|/|B|$).

Figure~\ref{toymodel} \emph{(left)} simulates for each user $50$ rounds of picking a video according to the model ($x$-axis represents time in number of videos picked). 
Each user (on $y$-axis) corresponds to a  horizontal line, with the color indicating the probability $p_B$ to get recommended a video from $B$. For instance a blue color indicates that the user on the $y$-axis has her history not containing a single video from $B$, which turns out to provoke a probability $p_B$ of 0 to get recommended a video from $B$.
Initially, some users have $p_B$ values different than $0$ or  $1$, while a significant number ($31$) of users already start with a $p_B$ equal to zero\footnote{Which meets the theoretical expectation: the probability of not picking a $B$ video is $(1-|V|/|B|)$, hence the probability of not picking a $B$ video $h$ times is $(1-|V|/|B|)^h\simeq 35\%$ chances given the selected parameters.}. 
As the time progresses, more and more users fall in either one of the following states: $p_B=0$, users never get recommended any $B$-type video, and $p_B=1$, users always get recommended $B$-type videos. 
In other words, users for which $p_B=1$ are trapped in the $B$-type RH. Before being locked in either state ($p_B\in \{0,1\}$, users in  the transient states are recommended both $A$-type and $B$-type videos. Although the figure shows only the first $t\leq 50$ rounds, the system ultimately converges to a state where all users are either only recommended $B$-type videos, either never recommended $B$-type videos. 

Of course, on real platform, users would not get recommended only videos from solely two categories, even when converged (this objective is often referred to as the need for \textit{serendipity} \cite{KOTKOV2016180}). 
This model can be trivially extended to adapt the specifics of particular setups. For instance, while we here focus on the analysis of a single RH ($B$), results are consistent in the case of multiple RHs (as confirmed empirically by simulations we made). 
On the analytical side, this model can be treated as an absorbing Markov Chain (which states capture all possible history states) to leverage the powerful analysis toolbox available in this domain.
User might as well choose videos on their own. However, while this model definitely constitutes an extreme case, we believe it is sufficient to illustrate the natural phenomenon behind the trapping dynamics into RHs: if a recommender suggests videos similar to past watched videos, and if users watch these videos, such a feedback loop naturally emerges. These feedback loops turn tight categories (either $A$ or $B$) into attractors, trapping users into a RH.

\begin{mdframed}[backgroundcolor=lgray]
\textbf{Summary:}  
We model a recommender's feedback loop: $B$-type videos are recommended to a user proportionally to this user's past consumption of $B$, while this user follows these recommendations. Simulations illustrate the trapping of users in a RH: the system quickly converges to a state where users either only either never see $B$ videos. The necessary ingredients are a recommender that users are interacting with by following recommendations, and attractor videos.
\end{mdframed}

\subsection{Detecting RHs from the user side}

Interestingly, this model also sheds light on black box observation scenarios (\ie from the perspective of users being recommended videos). We focus on one question: "can users detect they entered a RH?", to which we answer positively. 

\paragraph{Modeling user detection of RHs} After several videos have been consumed by every user, these users belong to three categories: users in $U_A$ never get recommended videos from $B$ ($p_B(u)=0$), users in $U_B$ always get recommended videos from $B$ ($p_B(u)=1$), and finally users in $U_{AB}$ that still get recommendations from both $A$ and $B$ categories.

Let $u,u'$ two users, and their collected recommendations $Y,Y'$. 
Since we defined $B$ to be a small subset of the whole video catalog, two $U_B$ users will likely have similar recommendations; in other words $Y\cap Y'$ is large. The same is also true for $U_A$ users, but to a lesser extent due to the larger set of videos in $A=V\setminus B$ and since $|V|>>|B|$. 
Finally, note that two users in different partitions will observe no intersection in their recommendations ($Y\cap Y'=\emptyset$)\footnote{Those observations also hold for the multiple RHs case: a collection of categories $\{B_i\}_i$, provided those are tight $\forall i,j, B_i\cap B_j=\emptyset$ and small compared to the overall catalog: $\forall i, |V|>>|B_i|$. In such setting, users in different RHs will also have an empty recommendation intersection.}.

We further analyze this model in Appendix \ref{appendix:user-sided-rh}, 
from which we extract the following results: given two users \textbf{in} the RH, their expected cosine similarity is $Sout=\frac{y}{|B|}$, and given two users \textbf{not} in the RH, their expected similarity is $Sin=\frac{y}{|V|-|B|}$, with $y=|Y|$ the number of recommendations returned by the platform. Since $B$-type videos only represent a small part of the whole video catalog $V$, those results confirm the intuition: users in the same RH will get a high similarity, hence this RH can be detected by exploiting the similarity of users recommendations. As this high similarity is synonym of a low diversity, we refer to this phenomenon as recommendation \emph{entropy collapse}.

\paragraph{Simulations} This is confirmed by the numerical simulations that are presented Figure \ref{toymodel} \textit{(right)}. It presents the evolution of the cosine similarity of four selected user couples. Users IDs $32$ and $16$ correspond to the two top dashed lines on the left plot, and correspond to users ending up in $U_B$ (that is, they both end in the RH). Their recommendation similarity, originally similar to the ones of others, increases sharply to around $y=0.5$. In contrast, users $5$ and $2$ correspond to the second pair of dashed lines on the left plot, are users ending up in $U_A$ (\ie not in the RH). Their recommendation similarity stays low and rather stable across time, at around $0.06$. Finally, the similarity between heterogeneous couples (namely $16-2$, $32-5$) drops to zero.

In other words, the recommendation similarity between two $U_B$ users will be high, whereas it will be low for two $U_A$ users, and zero between a $U_B$ and a $U_A$ user. Thus, by computing recommendations similarity, we can distinguish $U_B$ users from others and identify $B$ videos. 

\begin{mdframed}[backgroundcolor=lgray]
\textbf{Summary:} Our model captures the relation between users position with respect to a RH, and their recommendation vector similarity. This relation enables user-sided RH detection.
\end{mdframed}

\section{Collected data and the Rabbit-Hole effect}
\label{validation}

We now confront the data we collected with the proposed model, to conclude on the possibility to observe the RH effect with our automated approach.

\subsection{Down the RH through specific channel types?}
We choose to leverage the channel types identified by Ribeiro et al. in \cite{radicalization}.
They identified four types, namely the alt-right (AR), alt-light (AL), media (M) and the intellectual dark web (IDW).
We selected for each type the five most viewed channels that were still open at the time of our collection campaign (we for instance note that the most viewed channel for the AR category --named James Allsup-- was closed).
To these four, we added another arguably well separated type, the content providers for kids (K), we considered in the previous section. Table \ref{table:channels} lists the characteristics of these channels.
We stress that \textit{these labeled channels are used solely as a validation purpose} for our approach for detecting RHs, and that such external labeling is not required to perform the detection of RHs.

Each run by a bot proceeds as follows. A bot first randomly selects one channel among the ones we just listed. It then selects a video at random among all videos in that channel. This constitutes the starting video for its walk on the autoplay recommendations.

\paragraph{Experiment results}

\begin{figure}[h!]
\centerline{\includegraphics[width=\linewidth]{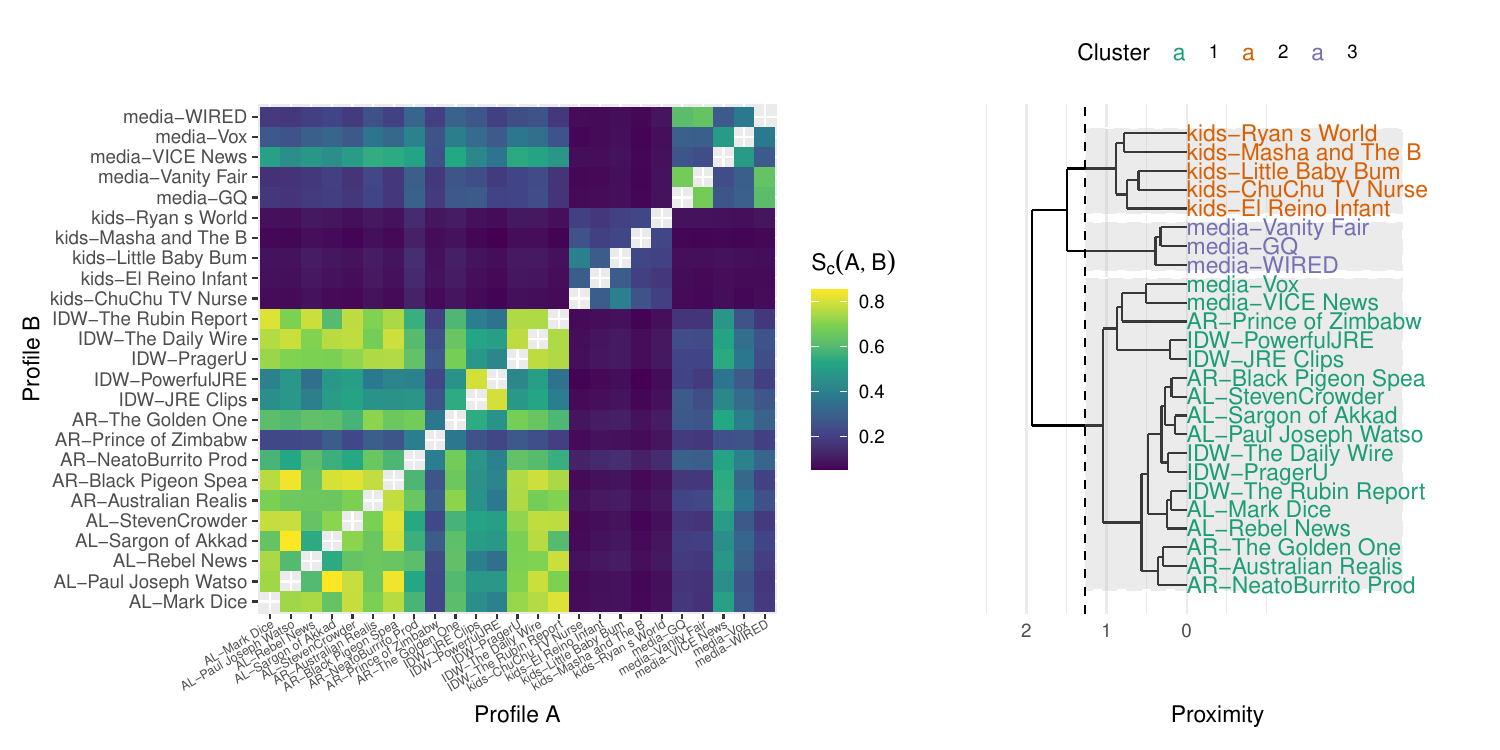}}
\caption{ (\textit{Left}) The similarity of recommendations after autoplay walks starting from specific channels, sorted by categories on both axes. Blocs appear, indicating the consistency of recommendations received by bots depending on their starting channels. (\textit{Right}) A dendrogram representation of the similarity between the channels, and the appearance of category-coherent clusters.}
\label{fig:proximity}
\end{figure}

We performed $5,438$ autoplay walks, each of length 5 from the various starting video channels. In Figure \ref{fig:proximity} are plotted the cosine similarities between the collected recommendation vectors at the end of each walk. It is based on the average blend of recommendations collected by each autoplay walk after its journey.
For instance a given walk ID on the $x$-axis is compared to another walk ID on the $y$-axis: the clearer the color the more similar are these two walks in the recommendations they receive (the diagonal is omitted for clarity: if $x=y$ then we have by construction a cosine similarity of 1). We grouped the channels by blocs following their types along both axes. 

A first important observation is the heatmap's overall bloc structure that closely matches the different investigated channel types. For instance, the media channels cause a bloc in the upper right part of the figure, which means that a walk starting from these media channels tends to get recommended videos that are similar to any other walk from starting from media channels, compared to any other walk starting from other channel types, say \eg kids channels. 

The inter bloc comparison reveals that the most dissimilar recommendations are made for the kids and AR, AL and IDW categories, as one can expect (this is to be observed by the dark cross appearing in the figure).
Similarities are high between the three AR, AL and IDW categories, as compared for instance to these three and the media category.

If we focus on the similarity within each bloc, we observe that walks starting from within the AR, AL and IDW channels are the closer in the recommendations they receive. This "internal" similarity is lower within the media bloc, and even lower within the kids bloc.
We note an outlier with the Prince of Zimbabwe channel, tagged as AR by \cite{radicalization}, yet appearing in our measurement closer to the media channels. Its last video was uploaded two years prior to our experiments.

The pairwise channel similarities of this heatmap can be exploited through a hierarchical clustering. Using Ward's method leads to the dendrogram presented in Figure \ref{fig:proximity} (\textit{right}). Cutting this dendrogram  reveals coherent clusters of channel categories. However, the threshold (proximity) one considers for cutting, symbolized as a dashed line on the Figure, leads to different clusters; this is studied in the next subsection. %

\subsection{Retrieving communities through clustering personalization}
\label{ss:clustering}

While Figure~\ref{fig:proximity} presents an inter-profile distance heatmap, it does not precisely define the boundaries of the potential RH into which walks are attracted. We now have a look at this aspect using clustering.

\begin{figure}[t!]
\centering
  \includegraphics[width=0.65\textwidth]{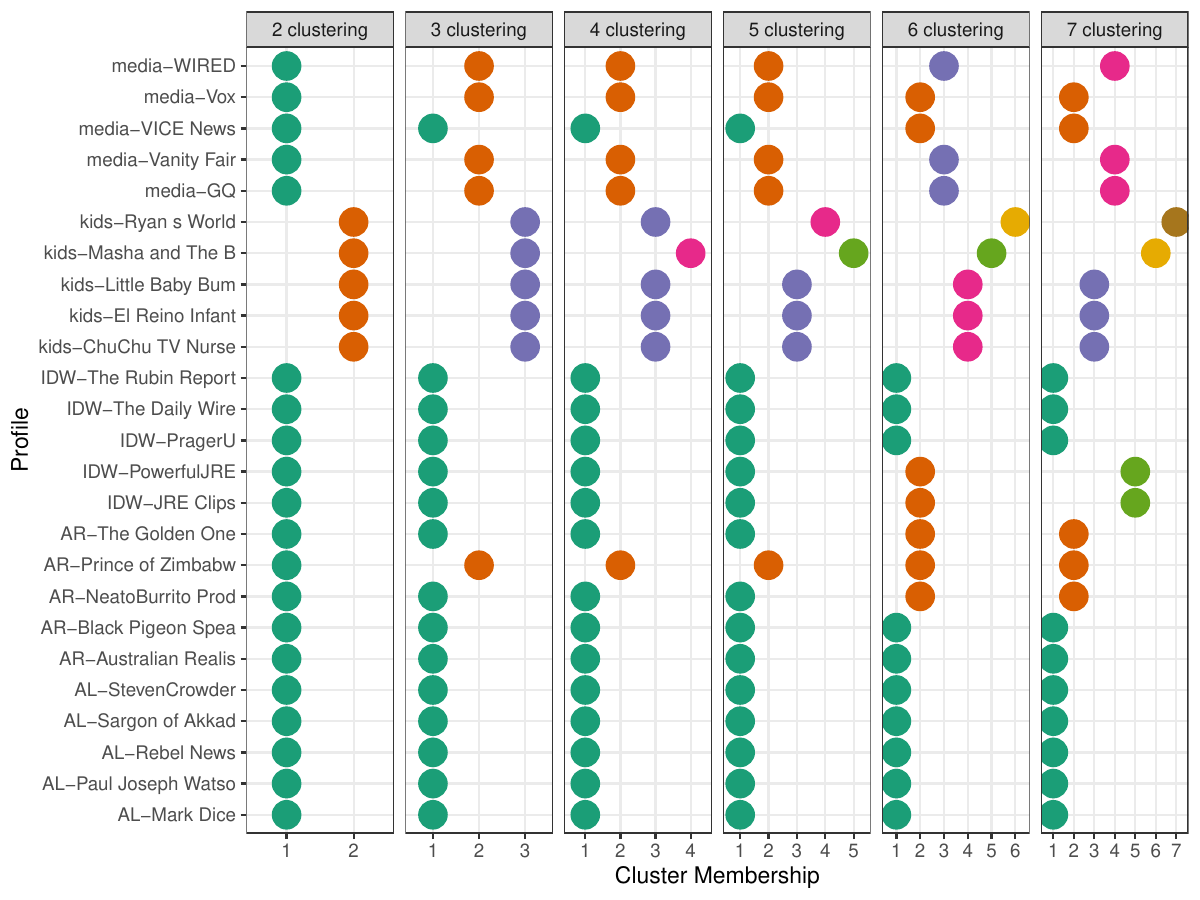}
\caption{Identifying rabbit-holes automatically using autoplay walks from channels on the $y$-axis: relying on $k$-Means. $k$ varies from 2 to 7, and resulting memberships ($x$-axis).}
\label{fig:clustering}
\end{figure}

We choose to focus on arguably the simplest clustering method, k-Means, which also has the advantage of being efficient to process the $16$ million recommendations collected for this experiment. 
k-Means is computed from the recommendation vectors described section \ref{botperso}; it requires as input $k$, the number of clusters to be found. 
Figure \ref{fig:clustering} presents the identified clusters memberships: each vignette represents the clustering results for a given value of $k$, ranging from $2$ to $7$, which allows to observe the various resolutions at which RHs could be defined. 

Partitioning the data into $k=2$ clusters splits the profiles into two broad categories: kids profiles (cluster 2-orange dots) and adult profiles (cluster 1-green dots). The $k=3$-clustering corresponds to the macro categories we defined for our investigations in Table \ref{table:channels}: kids, media, and right-wing profiles, with the notable exceptions of channel Prince of Zimbabwe that gets assigned to the media cluster, and Vice news assigned to the right-wing cluster. Further increasing $k$ refines the categories: $k=5$ splits the kids category, $k=6$ splits the right-wing category. 

\begin{table}[h!]
\centering
\footnotesize{
\begin{tabular}{rlrrrrrr}
  \hline
 & groups & 2 & 3 & 4 & 5 & 6 & 7 \\
  \hline
1 & BetweenSS/TotalSS & 0.20 & 0.31 & 0.41 & 0.49 & 0.56 & \textbf{0.62} \\
  2 & Rand: Original Labels & 0.52 & 0.71 & 0.70 & 0.69 & \textbf{0.77} & 0.76 \\
  3 & Rand:Right Merged & 0.76 & \textbf{0.88} & 0.87 & 0.86 & 0.77 & 0.76 \\
  4 & ARI: Original Labels & 0.18 & \textbf{0.32} & 0.27 & 0.24 & 0.27 & 0.23 \\ 
  5 & ARI: Right Merged & 0.53 & \textbf{0.75} & 0.72 & 0.70 & 0.47 & 0.45 \\ 
   \hline
\end{tabular} \caption{Clustering quality metrics for Figure \ref{fig:clustering}, as a function of the number of clusters created by the $k$-means.}
}
\label{table:clustering}
\end{table}


To complement these qualitative assessments regarding the identified RHs, one can rely on several metrics defined to measure the quality of a clustering. We choose to focus on three metrics; their evolution is depicted in Table 3,  
as a function of the number of identified clusters. The first one is a standard unsupervised metric that is based on the ratio between intra-cluster variance and inter-cluster variance, the intuition being that an ideal clustering is composed of very homogeneous clusters (low internal variance), that are all very different from each other (high inter-cluster variance). This measure steadily increases as the number of clusters grows -- as expected since larger $k$ values provide more freedom for $k$-Means to optimize. 

In addition to this unsupervised metric, we can also leverage the
categories assigned in \cite{radicalization}, as well as our kid
category, and use a supervised metric to measure to which extend the
clusters match these categories. We rely for this on the Rand index
\cite{rand}, that captures the percentage of correct assignments
issued by the clustering. It is maximized for $k=6$ clusters, reaching
a $77\%$ agreement. Interestingly, if one chooses to merge all
right-wing categories into a single category, one obtains a maximum at
$k=3$ with $88\%$ accuracy, which fits well the three broad studied
categories (media, right-wing and kids).
The Adjusted version of the Rand Index, noted ARI in the Table, also tops at $k=3$ with $75\%$ accuracy, confirming the previous statement.
Moreover, this shows how
this clustering captures with a reasonable accuracy the RHs tracked
statically and without personalization by the work in
\cite{radicalization}, although our findings rather indicate the
presence of a larger RH spanning the whole right-wing (AL, AR and IDW
categories).

\subsection{Measuring video attraction strength}
\label{hltext}

In this last experiment, rather than seeking the characterization of RH, we seek to characterize the strength of the video or category attractors.
In other words, instead of characterizing the destinations of the personalization made to bots as in the previous experiment, we measure the speed at which bots leave the mainstream video recommendations. 
The benefits of this perspective is to circumvent scale problems inherent to the notion of RH (how large is a RH? if $30\%$ of my recommendations are not in a RH, am I experiencing a RH?) and profit from a well defined reference we name \emph{mainstream recommendations}: recommendations made initially to the bots before they take any action. The intuition is that welcome page recommendations provided to new users without history ought to be somehow generic.

Concretely, given a set of profiles from which we perform multiple autoplay walks of depth $k$, we perform the following steps. As a first step, we consider only \emph{initial recommendation vectors}, collected by bots using fresh cookies before watching any videos. We compute the average initial recommendation vector $\hat{Y_\emptyset}$, which can informally be considered as the "mainstream barycenter". Then we observe the distribution of similarities between any initial recommendation vector and this center $\hat{Y_\emptyset}$. This allows to define a similarity threshold $\sigma$ above which a recommendation vector is considered as \emph{in the mainstream}. In other words, we train a classifier to detect recommendation vectors in the mainstream. Appendix Section~\ref{appendix:halflife-annex}
details the finding of this $\sigma$ ceil. 
Then, given any recommendation vector $Y$, if  $S_c(Y,\hat{Y_\emptyset})<\sigma$, we consider that the walk corresponding to $Y$ has left the mainstream. In addition, for a given profile that is the set of multiple walks, we are then able to compute the proportion of such walks that left the mainstream.

\begin{figure}[h!]
  \centering
\includegraphics[width=.65\textwidth]{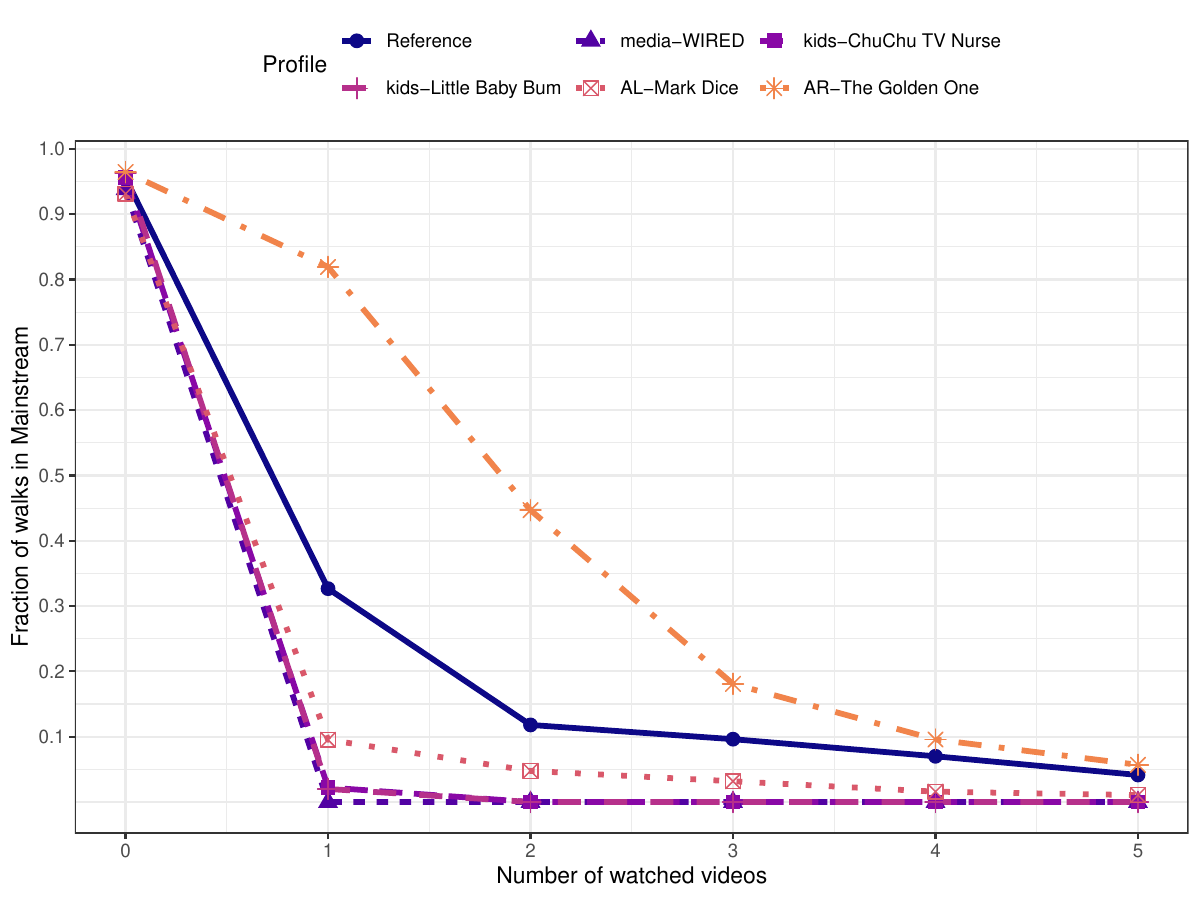}
\caption{Attraction strength of several starting videos (from some K, M, AL and AR channels from Table \ref{table:channels}).}
\label{fig:halflife}
\end{figure}

Figure \ref{fig:halflife} presents the collected data with this
perspective. It considers autoplay walks starting from several video
channels presented in Table \ref{table:channels}, and plots the
fraction of these that left the mainstream as a function of the number
of videos watched. For reference, we add a special profile coined
"Reference", that corresponds to autoplay walks issued from videos
picked randomly on the welcome page of a fresh bot. 

We stress that we only control the start video, as autoplay embed the recommender system's logic, so as soon as the bot accesses the recommended video, the walk can quit --or not-- the start category. The purpose of the experiment is thus to measure the  attraction strength of the starting video, and to observe how the rest of the walk performs with regards to the awaited loss of entropy.

Interestingly, autoplay walks starting at a kid video instantly leave the mainstream (\ie fractions of walks remaining in the mainstream is close to zero after the first video watched), so do the walks from the "media- Wired" channel. In comparison, only around 45$\%$ of the walks starting at "AR- The Golden One" channel videos left the mainstream after two videos watched, indicating that the attraction of these is much lower than the kids one, and also much lower than our reference walks. We conjecture that this is due to the recommendation algorithm possibly avoiding the autoplay proposition to extreme content systematically, leading walks to enter more classic recommendations in the followup. This would be consistent with the remark made in \cite{radicalization} noticing that \textit{more than $75\%$ of outgoing edges from all communities pointed towards the Other node}, that are videos outside the tracked radical channels.

\begin{figure}[h!]
\centerline{\includegraphics[width=0.8\linewidth]{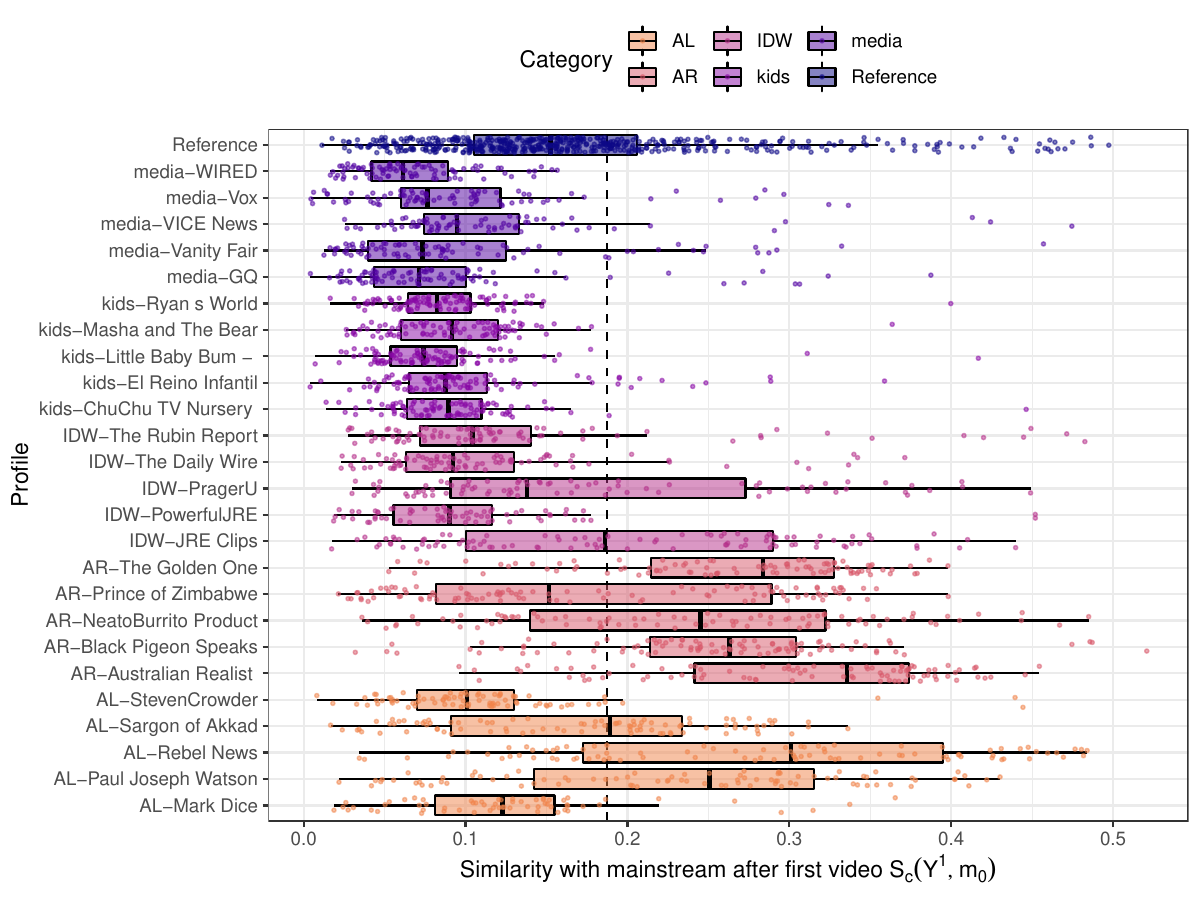}}
\caption{Mainstream similarity of each profile in all five considered categories (and control experiments starting on the welcome page, shown for for reference), after watching the first video. The vertical line represents the ceil selected in Appendix \ref{appendix:halflife-annex} and that defines the mainstream. For each profile, each individual walk corresponds to a point. In addition, a box plot represents the median and quartiles of the similarity distribution.}
\label{fig:hl-point3}
\end{figure}

\paragraph{A detailed view per channel}
For completeness, we now present Figure \ref{fig:hl-point3} as a display of the situation of every channel with regards to this attraction metric.
We observe the distribution of individual experiments aggregated in Figure \ref{fig:halflife}. 
Each point in a row represents the mainstream similarity of the recommendation vector extracted after watching a single video from a given walk, starting at a video of the channel on the $y$-axis. 
We focus on this first hop as our objective is to measure the mainstream similarity shift induced by a watching a video of each given channel. Since the subsequent videos watched in each walk are the suggested autoplay videos, they can be of different types depending on this initial personalization whose contribution to mainstream similarity becomes then harder to impute.
\\
We observe a distribution mass for each category that is distributed differently with regards to its similarity to the mainstream.  These results confirm what was observed in Figure \ref{fig:halflife}, in particular that starting a walk on an AR video tends to trigger recommendations leaving the bots in the mainstream. This is in stark contrast with the media or kids videos, that exhibit strong attraction.

While for all categories, the results are well separated by the threshold, we observe a high variance of the similarity scores for the channels in the AL category. This stresses that this metric can more precisely help to tag specific channels than what was arbitrarily used by Ribeiro et al. \cite{radicalization}. 

\begin{mdframed}[backgroundcolor=lgray]
\textbf{Summary:} Experiment results indicate that: \textit{i)} The automated autoplay walks from well separated video channels end up reproducing a RH effect, as indicated by the model. Walks from pre-labeled channels by a previous research paper consistently end up being recommended the same items.
\textit{ii)} Clustering the autoplay walks allows to retrieve communities with high recommendation similarities. This showcases the use of the approach to automatically retrieve various communities implied in a RH effect.
\textit{iii)} The attractivity of videos varies drastically, sometimes counter intuitively (\eg kids videos are stronger attractors than AR or AL videos in our collected data).
\end{mdframed}

\section{Limitations and Discussion} 

\paragraph{Simulating user behavior}
We have automated some actions of users on YouTube, following two main scenarios: the multiple consumption of videos of a single channel (Section \ref{ss:growing}), or consumption of the Autoplay videos recommended by YouTube (Section \ref{validation}). While we have shown that such an automation already allows for quantifying the RH effect, we believe that more advanced bot behaviors are still desirable (which quality of imitation can be measured through metrics \cite{bots}). Indeed, even better simulating humans could trigger different personalization behaviors overlooked by our approach. Yet, defining, implementing and analyzing the results produced by such human-realistic bot behaviors is challenging. We believe this increased complexity would conflict with the clarity, objectivity and simplicity required for the definition of a 'standard' measure as pursued in this paper.

\paragraph{A threshold to the RH}
While our bot-driven study permits to automatically retrieve video categories that were before identified by manual labeling (Section \ref{ss:clustering}), there remains a question of some thresholds to set. This is the case for the clustering performed in Figures \ref{fig:proximity} (right) and \ref{fig:clustering}, and for the threshold to consider in Figure \ref{fig:halflife}. An expert analysis of the results by our method is required to discriminate stable clusters in the former cases (for choosing a fix $k$), and to fix a threshold in the latter (as detailed in Appendix Section \ref{appendix:halflife-annex}).
We nevertheless believe these steps to be lightweight as compared to manual labeling of possibly millions of videos or channels. In addition, the setting of a threshold to decide when a user is entering a RH can be the prerogative of future auditors or regulators.

\paragraph{Are RHs a technical artifact?}
While the general opinion often tends to focus on specific and extreme topics for the observation of RHs (\eg political radicalization \cite{radicalization,ledwich2019algorithmic}), we have seen that our definition allows for the consideration of attractors videos or channels that are part of any category. Their attraction strength (Figure \ref{fig:halflife}) is to be evaluated without a priori, possibly yielding counter intuitive remarks: the alt-right (AR) videos cause a lesser entropy collapse of recommendations than kids ones. Our method is thus neutral when it comes to auditing, and able to measure from a user perspective the current dynamics of the formation of RHs around any target topic. We believe that those quantitative approaches, instead of competing against the necessary qualitative and/or subjective approaches required to study RH-related phenomena, rather provide an objective standpoint than can support those studies.

\section{Conclusion}
We presented in this paper an automated way to audit in vivo the personalization made to users by the YouTube recommender, with a focus on the RH effect.
We have shown that bots actually trigger personalization. 
We proposed a model that captures the  trapping dynamics of users in RHs. As a control experiment, we have shown that such an automated audit, without relying on any manual collection nor tagging, permits to extract video categories that have been studied manually by the previous state of the art paper on the topic.
This proposal is generic and a priori applicable to all recommenders, as it solely proceeds using the \texttt{url} of recommended items, and no other specific information. We believe this is an important step in the automated and continuous audit of recommender systems.

\appendix
\appendixpage

\section{An overview of the instrumentation in this study}
\label{appendix:instrumentation}

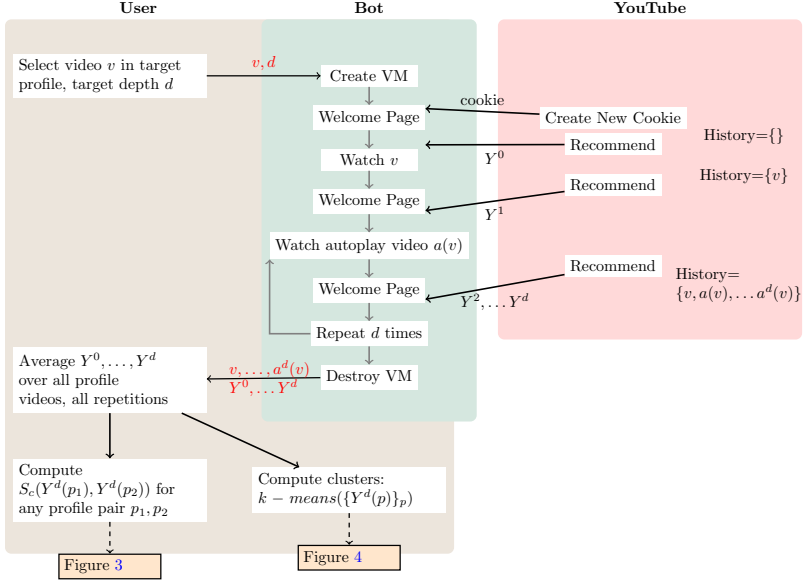
\begin{figure}[h!]
\scalebox{0.6}{
\pgfdeclarelayer{bg}    
\pgfsetlayers{bg,main}  
\definecolor{socC}{RGB}{116,174,150}
\definecolor{compC2}{RGB}{128,83,21}

\begin{tikzpicture}[align=left,node distance=.4cm]  
    \tikzstyle{block} = [draw, fill=blue!5, rectangle,     minimum
  height=3em, minimum width=6em]
\tikzstyle{shade} = [ rectangle,  minimum height=3em, minimum
width=6em,inner ysep=2em,inner xsep=0.5em, rounded corners]
  \tikzstyle{codestep} = [fill=white, rectangle,     minimum
  height=1em, minimum width=6em]
  \tikzstyle{hist} = [     minimum
  height=1em, minimum width=6em]
  \tikzstyle{userstep} = [fill=white, rectangle,     minimum
  height=1em, minimum width=6em,text width=4cm]

    \tikzstyle{output} = [fill=orange!20, draw, thick,rectangle,     minimum
  height=1em, minimum width=6em,text width=2cm]

  \node[codestep] (B)  {Create VM};
  \coordinate[above=.3cm of B] (botTop) ;
  \node[codestep] (C) [below= of B]  {Welcome Page};
  \node[codestep] (F) [below= of C]  {Watch $v$};
  \node[codestep] (G) [below= of F]  {Welcome Page};
  \node[codestep] (F1) [below= of G]  {Watch autoplay video $a(v)$};
  \node[codestep] (G1) [below= of F1]  {Welcome Page};
  \node[codestep] (H) [below= of G1]  {Repeat $d$ times};
  \node[codestep] (I) [below= of H]  {Destroy VM};

  \draw[line width=1pt, ->,gray] (H.west) -| (F1.south west); 

\foreach \source / \dest in
    {B/C,C/F,F/G,G/F1,F1/G1,G1/H,H/I}
  {
    \draw[line width=1pt,gray,->] (\source) -- (\dest);
  }

  \node[codestep] (Y1) [right=2.5cm of C]  {Create New Cookie};
  \coordinate[right=3cm of botTop] (ytTop) ;
  \node[codestep] (Y2) [below=.1cm of Y1]  { Recommend};
  \node[codestep] (Y3) [below= .4cm of Y2]  { Recommend};
  \node[codestep] (Y4) [below= 1.3cm of Y3]  { Recommend};

  \node[hist] (Hist1) [right=.2cm of Y1,yshift=-4mm] {History=$\{\}$};
  \node[hist] (Hist2) [below= .3cm of Hist1] {History=$\{v\}$};
  \node[hist] (Hist3) [below= 1.6cm of Hist2,xshift=-5mm,text width=2cm] {History=\\$\{v,a(v),\ldots a^d(v)\}$};

  \draw[line width=1pt, ->] (Y1) to node[above] {cookie} ([yshift=2mm] C.east);
  \draw[line width=1pt, ->] (Y2) to node[below] {$Y^0$} ([yshift=-6mm]  C.east);
    \draw[line width=1pt, ->] (Y3) to node[below] {$Y^1$}
    ([yshift=-2mm] G.east);
    \draw[line width=1pt, ->] (Y4) to node[below] {$Y^2,\ldots Y^d$} ([yshift=-2mm] G1.east); 


    \coordinate[left=5cm of botTop] (userTop) ;
    \node[userstep] (u1) [left=2.5cm of B] {Select video $v$ in target
      profile, target depth $d$};

    \node[userstep] (u2) [below=5.5cm of u1] {Average $Y^0,\ldots,Y^d$
      over all profile videos, all repetitions};

    \node[userstep] (u3) [below=1cm of u2] {Compute $S_c(Y^d(p_1),Y^d(p_2))$ for any profile pair $p_1,p_2$};
    \node[userstep] (u4) [right=1cm of u3] {Compute clusters: \\$k-means(\{Y^d(p)\}_p)$ };

    \node[output](o1) [below=.2cm of u3,yshift=-.5cm] {Figure~\ref{fig:proximity}};
    \node[output](o2) [below=.3cm of u4,yshift=-.4cm] {Figure~\ref{fig:clustering}};
    \draw[ thick,->,dashed] (u3) to (o1);
    \draw[ thick,->,dashed] (u4) to (o2);
    \draw[ thick,->] (I) to node[red,text width=3cm,pos=.2]{$v,\ldots,a^d(v)$ \\$ Y^0,\ldots Y^d$} (u2);
    \draw[thick,->] (u1) to node[above,color=red] {$v,d$}(B) ;

    \draw[->, line width=1pt] (u2) to (u3);
    \draw[->, line width=1pt] (u2) to (u4);
    
  \begin{pgfonlayer}{bg}    
             \node[shade, fill=compC2!15,fit={ (userTop) (u1)  (u2) (u3) (u4)},label={[xshift=-2cm] \textsc{\textbf{User}}}] (us) {};
    \node[shade, fill=socC!30,fit={ (botTop) (B)(C)(F)(G)(F1)(G1)(H)(I)},label={\textsc{\textbf{Bot}}}] (society) {};
      \node[shade, fill=red!15,fit={ (ytTop) (Y1)(Y2)(Y3)(Y4)(Hist2) (Hist3) },label={\textsc{\textbf{YouTube}}}] (ytb) {};
    \end{pgfonlayer}    
\end{tikzpicture}
}
\caption{Overview of the different steps performed in this study: our bots actions in YouTube, and the followup experiments.}
\label{fig:instrumentation}
\end{figure}

We present in Figure \ref{fig:instrumentation} the instrumentation logic for data extraction in this paper.

\section{A model for the Rabbit-Hole Effect}
\label{appendix:model}

Let $R$ be a recommender over a video catalog $V$. Let $B\subset V$ a subset of that constitutes, say, a specific video category (\eg kids, right-wing, portuguese language). We note $A=V \setminus B$. Suppose $R$ relies on a single feature in user profile "probability to recommend $B$ video" $p_B$ that is evaluated as the number of $B$ videos in a history $H$ of size $h$. Let $b=\vert B\vert,y=\vert Y\vert,v=\vert V\vert$. We assume users always pick a random video in recommendations $Y$. 

\paragraph{Clustering for automated identification of RHs}
Concretely, to audit a set of videos $T=t_1\ldots t_k$, we generate a set of fresh users $U$, whose first action will be to watch a video from $A$. We then let those users enter autoplay walks. After this walk, we collect the recommendations associated with this profile $\mathcal{Y}=\{Y(u)\}_{u\in U}$. Finally, we partition this set $\mathcal{Y}$ into a collection $S$ of $k$ clusters: $S=\{s_1,\ldots s_k\}$. While any clustering method might be used, one can leverage $k$-Means, which optimises: 
$ {arg\,min}_\mathbf{S}  \sum_{i=1}^{k} \, \frac{1}{2 |S_i|} \, \sum_{Y,Y' \in S_i} \left\| Y-Y' \right\|^2$. Since $k$-Means relies on random (centroids) initialization, it is considered good practice to conduct multiple runs and select the best obtained results. We followed the commonly prescribed value of $25$ such restarts. 

\section{User-Sided Rabbit-Hole Identification}
\label{appendix:user-sided-rh}

To estimate cosine similarities $S_c(Y,Y')$ in our model, first recall that $S_c(Y,Y')=\frac{Y.Y'}{\vert\vert Y \vert\vert \vert\vert Y'\vert\vert}=\frac{\vert Y\cap Y'\vert}{\sqrt{y}.\sqrt{y}}$ since we deal with binary vectors in this simple model (\ie a given video appears at most once in a recommendation vector). Hence estimating the intersection of $Y$ and $Y'$ is sufficient to compute the recommendation similarity. Let $X=|Y\cap Y'|$ be the random variable capturing this value. Let us call $p_{u\mapsto u'}$ the probability that a video in $Y$ is also present in $Y'$: we have $X\sim \mathcal{B}(y,p_{u\mapsto u'})$ where $\mathcal{B}$ is the binomial law. We then split cases:
\begin{itemize}
    \item $u,u'\in U_B$. In this case $u$ has picked $y$ videos among the $b$ videos of $B$. Hence $p_{u\mapsto u'}=y/b$.
    \item $u,u' \in U_A$. In this case, videos are picked in $V\setminus B$. Hence $p_{u\mapsto u'}=y/(v-b)$.
\end{itemize}

Since the expected value of a binomial law $\mathcal{B}(n,p)$ is simply $np$, one can further develop $E[S_c(u,u')]$:
\begin{itemize}
    \item $u,u'\in U_B$. Since $p_{u\mapsto u'}=y/b$, $E[X]=y^2/b$, and thus $E[S_c(u,u')]=\frac{y}{b}$.
    \item $u,u' \in U_A$. Since $p_{u\mapsto u'}=y/(v-b)$, $E[X]=y^2/(v-b)$, and thus $E[S_c(u,u')]=\frac{y}{(v-b)}$.
\end{itemize}

When applied to the system instance detailed above ($v=1000,b=100,y=50$), one gets $E[S_c(u,u') | u,u'\in U_B]=1/2$, and $E[S_c(u,u') | u,u'\in U_A]=1/18$, which is validated by observations in Figure \ref{toymodel}.

\section{Estimating the Attraction Strength of Videos}
\label{appendix:halflife-annex}

We here extend the technical parts of Section~\ref{hltext}. 
Let $W$ a set of walks of length $k$, and let $\forall 0\leq i<k, \forall w\in W, Y^i(w)$ be the recommendation vector collected by walk $w$ after watching $i$ videos.

First, we compute the mainstream recommendation vector as the mean of all recommendation vectors observed from any walk before its first watched video: $m_\emptyset= 1/|W|\sum_{w\in W} Y^0(w)$.

\begin{figure}[h!]
\centerline{\includegraphics[width=0.65\linewidth]{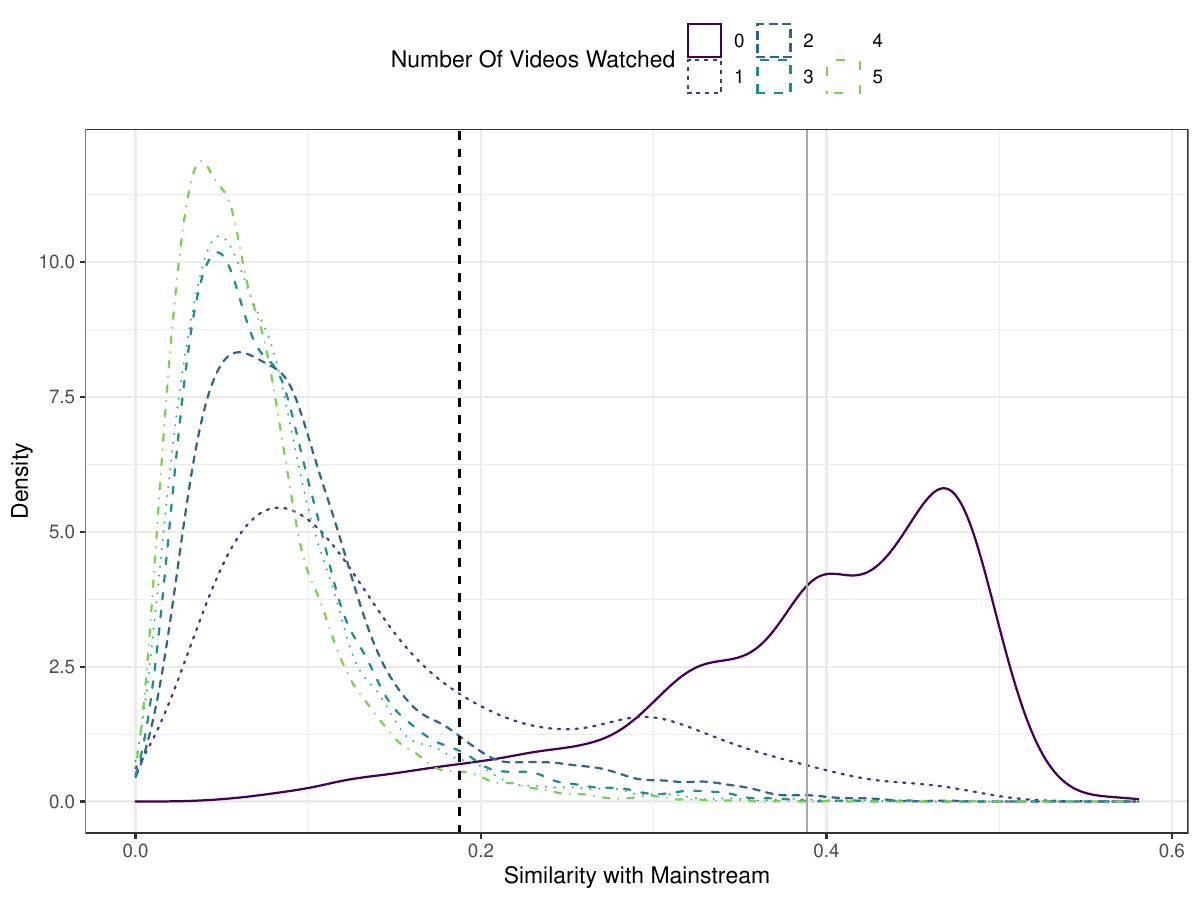}}
\caption{Distribution of cosine similarity against average mainstream recommendation, as a function of the number of watched videos.}
\label{fig:halflife-dens}
\end{figure}

Yet, recommendation vectors are noisy by nature, hence to qualify "mainstream recommendation", we need to understand how this noise perturbs observations. To do so, we look at the similarity between recommendation vectors in the mainstream by construction (\ie no interactions nor videos watched by bots) and pick a similarity threshold that includes the vast majority of those vectors in the mainstream. Figure~\ref{fig:halflife-dens} shows the density of this distribution, along with its evolution as videos are watched. We can observe a clear gap between $0$ video watched, and $i>1$ videos watched. The gray vertical line represents the average similarity observed over the dataset (0.389). The dashed line represents the selected ceil : $\sigma=$0.188. In other word, any walk $Y$ such that $S_c(Y,m_\emptyset)>\sigma$ is classified as "in the mainstream". This ceil $\sigma$ is set so that $95\%$ of the $Y^0$ vectors are in the mainstream.

\bibliography{references}

\end{document}